\documentclass[journal]{IEEEtran}

\usepackage[T1]{fontenc}
\usepackage{textcomp}

\usepackage{cite}

\usepackage{graphicx}
\usepackage[table]{xcolor}
\usepackage{flafter} 
\usepackage{placeins}

\usepackage{amsmath}
\usepackage{amssymb}
\usepackage{amsthm}
\usepackage{comment}
\usepackage{array}
\usepackage{tabularx}
\usepackage{subfigure}
\usepackage{booktabs}
\usepackage{multirow}
\newcommand{\papertablestyle}{%
  \normalfont\rmfamily\footnotesize
  \renewcommand{\arraystretch}{1.5}}
\newcommand{\tablehead}[1]{%
  {\renewcommand{\arraystretch}{1.05}%
   \begin{tabular}[c]{@{}c@{}}#1\end{tabular}}}

\usepackage{listings}
\usepackage{lstlinebgrd}
\usepackage[linesnumbered,ruled,lined]{algorithm2e}
\DontPrintSemicolon
\SetKw{KwContinue}{continue}
\SetAlgoCaptionSeparator{}
\SetAlCapSty{textbf}
\SetAlCapNameSty{textnormal}
\SetAlFnt{\normalfont\small}
\SetArgSty{textnormal}
\SetAlgoNlRelativeSize{-2}
\SetNlSkip{.65em}
\SetAlgoInsideSkip{smallskip}
\SetAlgoLined
\SetAlgoShortEnd
\SetKwInput{KwInput}{Input}
\SetKwInput{KwOutput}{Output}
\newcommand{\algcomment}[1]{\textcolor{paperlinkblue}{\textit{#1}}}
\newcommand{\algstage}[1]{\algcomment{\#\#\# #1 \#\#\#}}
\newcommand{\algfn}[1]{\ensuremath{f_{\mathrm{#1}}}}

\usepackage[normalem]{ulem}

\usepackage{url}

\definecolor{paperlinkblue}{rgb}{0.3,0.2,0.9}
\usepackage[colorlinks,linkcolor=paperlinkblue,anchorcolor=paperlinkblue,citecolor=paperlinkblue]{hyperref}
\hypersetup{pdftitle={LineageForge: Constructing Verifiable Smart Contract Vulnerability Datasets at Scale}}

\makeatletter
\let\paper@makecaption\@makecaption
\long\def\@makecaption#1#2{%
  \ifx\@captype\@IEEEtablestring
    \begingroup
      \par\@parboxrestore\normalfont\footnotesize
      \setbox\@tempboxa\hbox{#1:\nobreakspace #2}%
      \ifdim\wd\@tempboxa>\hsize\else\centering\fi
      \@IEEEtabletopskipstrut #1:\nobreakspace #2\par
      \addvspace{0.5\baselineskip}%
    \endgroup
    \@IEEEtablecaptionsepspace
  \else
    \paper@makecaption{#1}{#2}%
  \fi}
\makeatother

\newif\ifshowappendix
\showappendixfalse 
\newcommand{\tool}{LineageForge}

\theoremstyle{definition}

\lstdefinelanguage{Solidity}{
  keywords={contract,function,public,external,private,internal,view,pure,returns,
    return,require,if,else,for,while,address,uint,uint256,bool,string,memory,
    calldata,storage,modifier,event,emit,interface,library,import,pragma},
  keywordstyle=\color{blue!60!black}\bfseries,
  sensitive=true,
  comment=[l]{//},
  morecomment=[s]{/*}{*/},
  commentstyle=\color{green!40!black},
  morestring=[b]",
  stringstyle=\color{red!50!black}
}
\lstdefinestyle{paperSolidity}{
  language=Solidity,
  basicstyle=\ttfamily\footnotesize,
  columns=fullflexible,
  breaklines=true,
  breakatwhitespace=false,
  keepspaces=true,
  showstringspaces=false,
  frame=lines,
  captionpos=b,
  abovecaptionskip=0.4em,
  belowcaptionskip=0.6em,
  aboveskip=0.7em,
  belowskip=0.9em,
  numbers=left,
  numberstyle=\tiny\color{black!55},
  numbersep=5pt,
  xleftmargin=1.3em,
  framexleftmargin=0.3em,
  linewidth=\linewidth,
  escapeinside={(*@}{@*)}
}

\definecolor{tableHeaderOrange}{HTML}{F0D6B5}

\definecolor{tableRowOne}{HTML}{FFFBF6}
\definecolor{tableRowTwo}{HTML}{FDF6ED}
\definecolor{tableRowThree}{HTML}{FBF1E7}
\definecolor{tableRowFour}{HTML}{F9ECDF}
\definecolor{tableRowFive}{HTML}{F7E7D7}
\definecolor{tableRowSix}{HTML}{F5E2CE}
\definecolor{blue}{rgb}{0.3, 0.2, 0.9}
\definecolor{tableToolColumn}{HTML}{F5E7D8}
\definecolor{tableToolHeader}{HTML}{E8C393}
\definecolor{tableAccent}{HTML}{A85E28}
\definecolor{tableHeaderText}{HTML}{704321}

\begin{document}
\bstctlcite{paper_bib_control}
%
\title{\tool{}: Large Language Model for Constructing Verifiable Smart Contract\\Vulnerability Datasets at Scale}

\author{Yucheng~Huang, Jinbo~Wen, Jiawen~Kang, \textit{Senior Member, IEEE},\\
Tenghui~Huang, Ming~Li, Chengjun~Cai, Hongfa~Ding, and Shuyuan~Lin%

\thanks{%
Y.~Huang and S.~Lin are with the College of Cyber Security, Jinan University, Guangzhou, China (e-mails: yucheng.huanggd@gmail.com, swin.shuyuan.lin@gmail.com).
J.~Wen is with the Department of Computer Science, City University of Hong Kong, Hong Kong SAR, China (e-mail: jinbowen2-c@my.cityu.edu.hk).
J.~Kang and T.~Huang are with the School of Automation, Guangdong University of Technology, Guangzhou, China (e-mails: kavinkang@gdut.edu.cn, 3123000938@mail2.gdut.edu.cn).
M.~Li is with the Dingyi Pei Institute of Cryptography, Guangzhou University, Guangzhou, China (e-mail: limjnu@gmail.com).
C.~Cai is with the Department of Computer Science, City University of Hong Kong (Dongguan), Dongguan, China (e-mail: chengjun.cai@cityu-dg.edu.cn).
H.~Ding is with the School of Information, Guizhou University of Finance and Economics, Guiyang, China (e-mail: hongfa.ding@foxmail.com).

\textit{Corresponding author: Jiawen Kang.}
}
}



\maketitle

\begin{abstract}
Smart contract vulnerabilities pose serious security risks to decentralized applications. As deployed smart contract code is generally immutable, it is particularly important to detect vulnerabilities before deployment. To this end, Smart Contract Vulnerability (SCV) datasets play a key role in developing and evaluating vulnerability detection tools for securing smart contracts. Existing SCV datasets are commonly constructed through tool-based labeling, manual curation, or vulnerability injection based on predefined bug patterns. However, these approaches have three key challenges: First, vulnerability samples often lack traceable origins in real-world security incidents. Second, ground-truth labels often lack empirical validation. Third, the number of instances for individual vulnerability types remains limited even when both provenance and validation are available. To address these challenges, we present \tool{}, a framework that leverages Large Language Models (LLMs) to construct verifiable SCV datasets at scale from vulnerabilities exploited in real-world attacks. \tool{} consists of two modules: 1) The incident-guided discovery module identifies compatible contracts whose business logic supports a target vulnerability and collects supporting source evidence. 2) Guided by this evidence, the Evidence-Grounded Contract Construction (EGCC) module constructs a protected contract that is free of the target vulnerability and introduces targeted code changes to derive a corresponding vulnerable version. EGCC then adapts the attack to each contract and executes it against both the protected and vulnerable versions. 
Based on seven real-world incidents, \tool{} produces $1{,}059$ validated instances, each pairing a protected contract with its vulnerable counterpart and including executable proofs of concept and the execution environment for reproducible verification. \tool{} is available at \url{https://github.com/youfafa/lineageforge-dataset}.
\end{abstract}


\section{Introduction}
\label{sec:introduction}

Blockchain is a decentralized infrastructure to record transactions and execute smart contracts for Web~4.0~\cite{zhang2019securityprivacy,ec2023web4}. Smart contracts deployed on blockchains automatically execute predefined logic to manage digital assets and implement incentive mechanisms~\cite{wen2025lmmincentive}. However, attackers may manipulate contract execution and cause large financial losses by exploiting vulnerabilities in smart contracts~\cite{ivanov2023security}. For example, in the recent LegendaryMoneyMonNft incident, a signature-validation flaw enabled attackers to exploit the contract and cause approximately \$85.5K in losses\footnote{\href{https://github.com/SunWeb3Sec/DeFiHackLabs\#20260528-legendarymoneymonnft---ecrecover-address0-signature-bypass}{LegendaryMoneyMonNft incident (DeFiHackLabs)}}, which underscores the need for reliable smart contract vulnerability detection. Therefore, it is particularly important to train and evaluate detection tools on well-designed and large-scale Smart Contract Vulnerability (SCV) datasets~\cite{huang2026paravul}. Existing studies construct SCV datasets by three main approaches: automated vulnerability labeling, collection of real-world security evidence, and synthetic vulnerability generation. For automated labeling, SmartBugs provides a benchmark of vulnerable contracts, while DIVE aggregates results from multiple analysis tools to assign vulnerability labels~\cite{smartbugs2020,dive2026}. For the datasets based on real-world evidence, DAppSCAN and FORGE link audit findings to contract source code~\cite{dappscan2023,forge2026}. EVMbench converts audit findings into executable exploit tests with corresponding execution environments~\cite{evmbench2026}, while CyberChainBench organizes historical attacks into executable benchmarks~\cite{cyberchainbench2026}. For synthetic vulnerability generation, SolidiFI and MuSE inject bugs into existing contracts using predefined bug patterns or mutation operators~\cite{solidifi2020,muse2025}. XploGen instead generates contract variants that violate specified business rules and synthesizes corresponding attacks~\cite{xplogen2024}.

Despite these efforts, existing SCV datasets still face three key challenges: First, vulnerabilities introduced by using predefined mutation operators or fixed rules often lack traceable links to real-world incidents, making their practical relevance difficult to assess. Second, SCV datasets labeled by automated analysis tools often lack execution-based validation to establish ground truth. Tool-generated labels may contain false positives~\cite{chen2025smartguard}, and such labels alone cannot establish whether the reported vulnerabilities can be triggered in practice. Third, even when vulnerabilities have traceable real-world origins and their exploitability has been verified, the number of contract instances derived from a single source vulnerability remains limited, which limits the ability to evaluate whether a detector can recognize the same vulnerability across different contract implementations. The importance of such evaluation is illustrated in~\cite{solidifi2020}, which shows that a minor syntactic change preserving contract semantics causes SmartCheck to miss a previously detected timestamp-dependency vulnerability.

To address these challenges, we present \tool{}, a Large Language Model (LLM)-based framework for constructing verifiable SCV datasets at scale from vulnerabilities observed in real-world attacks. Before dataset construction, an agent-assisted preparation pipeline iteratively refines a lineage specification $\mathcal{L}$ based on real-world attack evidence and feedback from trial runs. The specification is finalized when the estimated number of compatible contracts is sufficient and the quality of the associated Proof-of-Concept (PoC) attacks meets predefined requirements. The resulting $\mathcal{L}$ characterizes the target vulnerability and defines the requirements for the Incident-Guided Discovery (IGD) and Evidence-Grounded Contract Construction (EGCC) modules. Given $\mathcal{L}$, IGD identifies source-verified contracts whose business logic supports the target vulnerability and prepares the necessary evidence and construction requirements. EGCC then guides LLMs to construct a protected contract and derive its vulnerable counterpart through targeted code changes. It validates each pair through PoC execution and source-code inspection, and packages accepted pairs with their PoCs and execution environments as dataset records. 
The main contributions of this paper are summarized as follows:

\begin{itemize}

  \item We present \tool{}, a novel LLM-based framework for constructing verifiable SCV datasets at scale from vulnerabilities observed in real-world attacks. Specifically, \tool{} applies each target vulnerability to compatible contracts from real deployments, which effectively extends vulnerability datasets beyond the limited set of contracts involved in the original incidents while preserving traceability to real-world vulnerability sources.

  \item We devise a new IGD module to identify a predefined number of compatible contracts from a source-verified contract corpus. IGD first screens contract interfaces and code features. It then analyzes source code under the guidance of the lineage specification $\mathcal{L}$ to identify contracts with the required business logic. For each candidate, IGD packages the source code, supporting evidence, and construction requirements for EGCC.

  \item We design an EGCC module to construct and validate vulnerability instances using $\mathcal{L}$ and the source packages produced by IGD. EGCC guides LLMs to construct a protected contract and derive its vulnerable counterpart through targeted code changes. It then adapts and executes equivalent PoC attacks against both contracts. Source-code inspection and attack outcomes are jointly examined to verify that the intended exploit succeeds on the vulnerable contract but fails on the protected one.

  \item By employing \tool{}, we construct $1{,}059$ validated dataset records from vulnerabilities associated with seven real-world incidents. Each record contains a protected contract, its vulnerable counterpart, executable PoC attacks, and the execution environment. Experiments show that training with corresponding protected contracts as negative examples achieves higher pairwise accuracy and Macro-F1 than training with unrelated negative examples for the same vulnerable contracts.

\end{itemize}

The remainder of this paper is organized as follows. Section~\ref{sec:related} reviews related work. Section~\ref{sec:framework} introduces the architecture of \tool{}. Sections~\ref{sec:discovery} and~\ref{sec:construction} introduce the IGD and EGCC modules, respectively. Section~\ref{sec:evaluation} presents the evaluation results, and Section~\ref{sec:conclusion} concludes the paper.

\section{Related Work}
\label{sec:related}

\subsection{Vulnerability Datasets and Benchmarks}

Existing SCV datasets link contract source code with vulnerability annotations to support the development and evaluation of vulnerability detectors. For instance, SmartBugs provided collections of contracts and a framework for running and comparing analysis tools, while SmartBugs 2.0 offered a unified interface for tool integration~\cite{smartbugs2020,durieux2020empirical,diangelo2023smartbugs2}. DIVE combined multi-tool voting with rule-based source-code checks to generate multi-label vulnerability annotations~\cite{dive2026}. ACToolBench curated access-control vulnerabilities from public disclosures, attack incidents, and audit reports~\cite{actoolbench2025}. DAppSCAN and FORGE associated audit findings with contract source code via manual curation and LLM-based extraction, respectively~\cite{dappscan2023,forge2026}. However, these SCV datasets generally lacked executable PoCs and corresponding execution environments to establish whether the annotated vulnerabilities could be triggered in practice.

Reliable ground truth requires validating reported vulnerabilities via attack execution against the corresponding contracts. This is important because tool-reported vulnerabilities may not be exploitable in practice. For example, Hu et al.~\cite{hu2025unexploited} found that many contracts flagged by the detectors they examined could not be exploited. To support execution-based verification, CyberChainBench organized historical attacks into exploitation and patching benchmarks~\cite{cyberchainbench2026}, while EVMbench provided detection, patching, and exploitation tasks based primarily on audit findings, along with manually prepared exploit tests and execution environments~\cite{evmbench2026}. Despite these advances, the number of execution-verified contract instances derived from the same source vulnerability remained limited, restricting evaluation of whether detectors could recognize the same vulnerability across different contract implementations.

\subsection{Vulnerability Injection and Synthesis}

Vulnerability injection and synthesis expand datasets beyond collected cases by generating samples from existing contracts. Early studies investigated mutation testing in software engineering. Petrovi\'c et al.~\cite{petrovic2022mutation} studied mutation testing at Google, G\"orz et al.~\cite{gorz2023mutation} developed a mutation-analysis benchmark for fuzzers, and S\'anchez et al.~\cite{sanchez2024mutation} examined mutation-testing practices in open-source projects. In the smart contract domain, Deviant~\cite{deviant2019} and SuMo~\cite{sumo2022} employed Solidity-specific mutation operators to evaluate the fault-detection capabilities of test suites. SolidiFI~\cite{solidifi2020} and MuSE~\cite{muse2025} injected predefined vulnerability patterns into existing contracts, while SCAnoGenerator used control-flow and data-flow analysis to guide anomaly injection~\cite{scanogenerator2024}. Moving beyond predefined operators, XploGen leveraged LLMs to generate vulnerable contract variants and corresponding attacks from business-logic specifications, checking for specification violations during execution~\cite{xplogen2024}. Despite these advances, injecting predefined vulnerability patterns or violating specified business rules does not establish that a generated vulnerability corresponds to one exploited in a real-world attack. Without traceable links to documented incidents, generated samples may not reflect security risks that arise in practice. This motivates approaches that derive vulnerabilities from real-world incidents and verify whether the same vulnerability can be reproduced across different contract implementations.

\begin{figure*}[t]
\centering
\includegraphics[width=\textwidth]{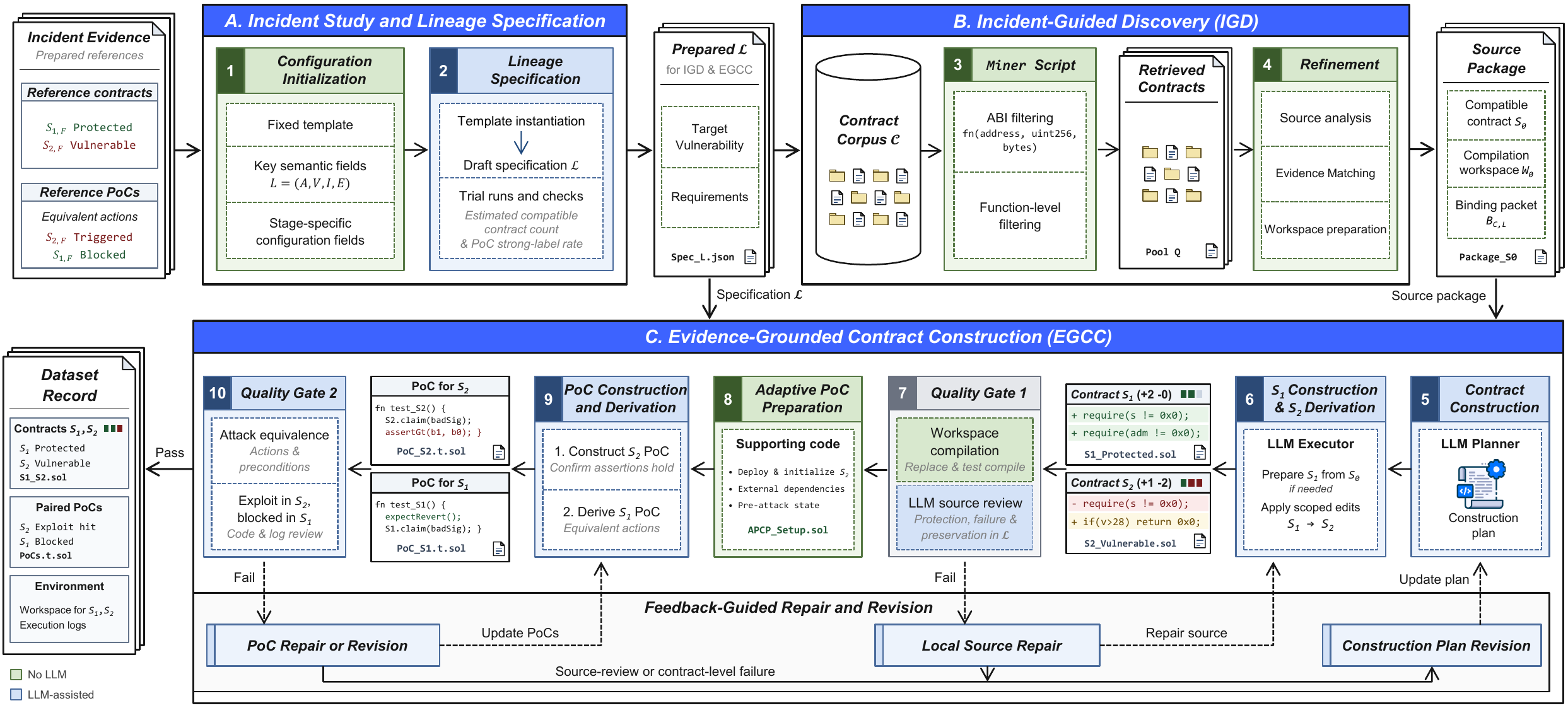}
\caption{Overview of \tool{}. (A) The lineage specification is prepared based on $S_{1,F}$, $S_{2,F}$, and their corresponding PoCs. (B) IGD identifies compatible contracts and passes them to EGCC along with their compilation workspaces and binding packets. (C) EGCC constructs $S_1$ and $S_2$, and executes the corresponding PoCs. Blue boxes denote LLM-assisted operations, while green boxes denote operations performed without LLM assistance.}
\label{fig:pipeline}
\end{figure*}

\subsection{Exploit Generation and Adaptation}

Exploit generation provides executable evidence of whether a vulnerability can be triggered in practice. For instance, ExGen generated attacks through symbolic execution~\cite{jin2023exgen}, while SmarTest guided symbolic execution using a statistical language model~\cite{so2021smartest}. Fuzzing-based approaches, including Harvey~\cite{wuestholz2020harvey}, sFuzz~\cite{nguyen2020sfuzz}, SMARTIAN~\cite{choi2021smartian}, ILF~\cite{he2019ilf}, ContraMaster~\cite{wang2022contramaster}, ItyFuzz~\cite{shou2023ityfuzz}, and EF/CF~\cite{rodler2023efcf}, explored transaction inputs and execution states to expose vulnerabilities. Several studies further guided this exploration: xFuzz applied machine learning to cross-contract fuzzing~\cite{xue2024xfuzz}, IR-Fuzz prioritized function invocations and important branches~\cite{liu2023irfuzz}, and Vulseye directed testing toward selected code and state targets~\cite{liang2025vulseye}. AdvSCanner and PoCo leveraged LLMs to generate reentrancy attacks and audit-guided Foundry exploits, respectively~\cite{advscanner2024,poco2025}. V2E evaluated vulnerability triggerability and attack profitability~\cite{v2e2026}, while KASS combined retrieval-based planning with execution feedback~\cite{kass2026}. Attack execution was also used to evaluate patches by comparing contract behavior before and after repair~\cite{bobadilla2026}. Beyond generating attacks for individual contracts, exploit adaptation reuses historical attacks against related contracts. PoCShift combined code-clone matching with property-based feasibility analysis to adapt historical PoCs to contracts with similar vulnerable code and verify their effects~\cite{pocshift2025}. Building on historical attack reuse, \tool{} identifies contracts whose business logic supports a vulnerability observed in a real-world incident. 

\section{The Architecture of LineageForge}
\label{sec:framework}

\subsection{Task Definition of LineageForge}
\label{sec:definitions}

Given a vulnerability observed in a selected real-world incident and a source-verified smart contract corpus, the task is to identify compatible contracts and construct verifiable SCV datasets at scale. For each compatible contract, \tool{} first constructs a contract $S_1$ protected against the target vulnerability and derives a corresponding vulnerable contract $S_2$ through targeted code changes. It then executes corresponding PoC attacks against both contracts under equivalent attacker actions. A contract pair is accepted only if the attack succeeds on $S_2$ but fails on $S_1$, providing evidence that the targeted code changes introduce the intended vulnerability. Each resulting dataset record contains the verified contract pair, the associated PoC attacks, and their execution environments. To balance broad applicability with the feasibility of contract adaptation and execution-based verification, we focus on vulnerabilities whose security conditions and affected business operations can be attributed to a single contract and whose exploitation can be triggered through a public or external function.

\subsection{Lineage Specification Preparation}
\label{sec:incident-preparation}

\begin{figure}[t]
\centering
\includegraphics[width=\columnwidth]{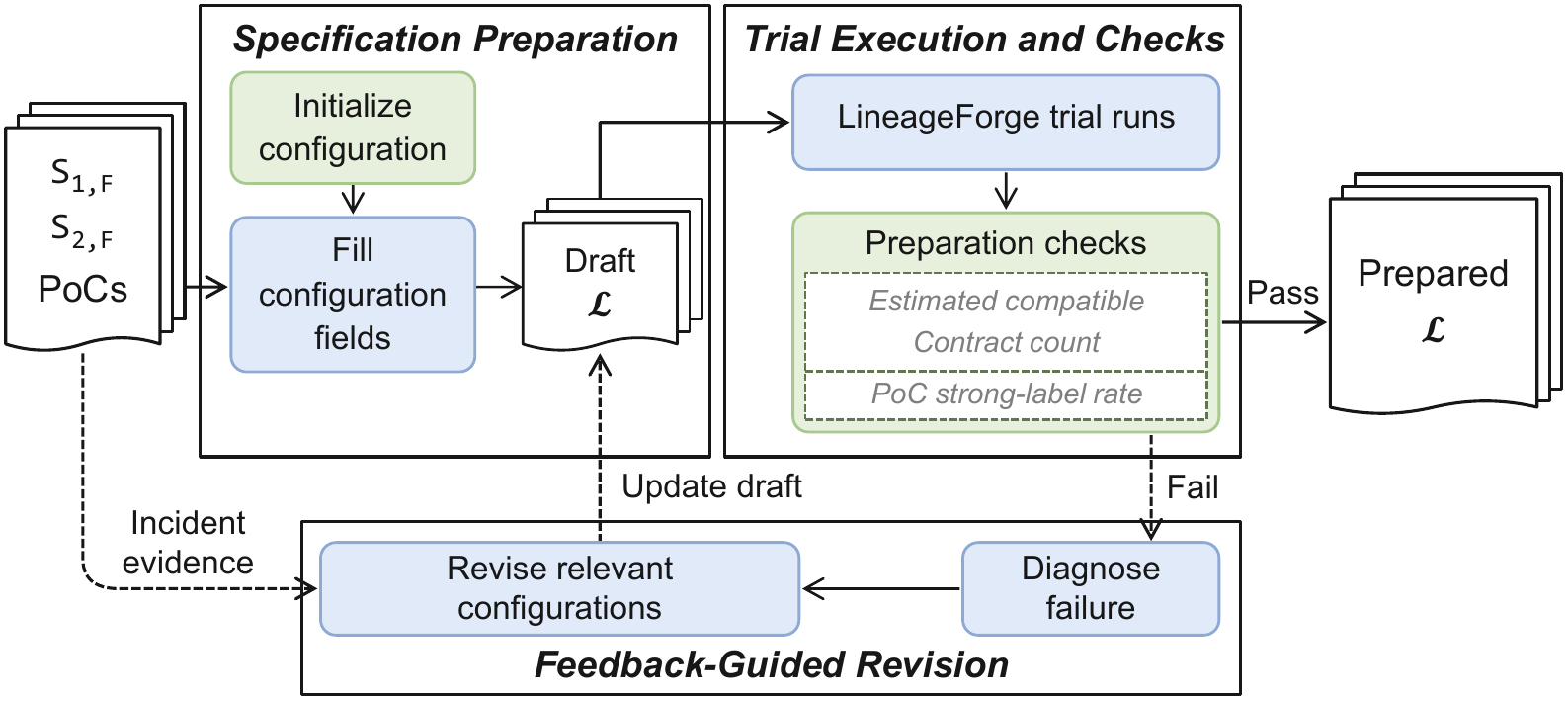}
\caption{Workflow for lineage specification preparation.}
\label{fig:specification-preparation}
\end{figure}

As shown in Fig.~\ref{fig:pipeline}A, we use evidence from the selected incident to construct a protected reference contract $S_{1,F}$ and its vulnerable counterpart $S_{2,F}$ in a local environment. We then execute PoCs derived from the historical attack against both contracts under equivalent attacker actions. The execution results confirm that the attack succeeds on $S_{2,F}$ but is blocked by the protection in $S_{1,F}$. Based on $S_{1,F}$, $S_{2,F}$, and their corresponding PoCs, we characterize the vulnerability lineage using a semantic tuple $L=(A,V,I,E)$. Here, $A$ denotes attacker-controlled inputs or roles, $V$ denotes the relevant subject or authority, $I$ denotes the required security condition, and $E$ denotes the protected business operation or state change governed by $I$. The target effect is the outcome resulting from a violation of $I$, such as an unauthorized payout. Based on $L$, we define the lineage specification $\mathcal{L}$, which describes the target vulnerability and specifies the requirements for each stage of \tool{}. These requirements cover the business logic and source evidence needed for contract discovery, and the expected protection in $S_1$, vulnerability in $S_2$, and corresponding PoC execution outcomes.

As shown in Fig.~\ref{fig:specification-preparation}, we initialize a new lineage specification $\mathcal{L}$ using a fixed template that defines its required fields. We first populate the template using the source code of $S_{1,F}$ and $S_{2,F}$ and their corresponding PoCs to generate a draft specification. We then conduct trial runs of \tool{} and evaluate two preparation criteria. First, the estimated number of compatible contracts in the full corpus must meet a preset threshold. Second, the strong-label rate among trial samples must satisfy its corresponding threshold. The strong-label rate is the proportion of trial samples assigned the highest-quality label by the PoC reviewer, an LLM-based component that assesses PoC quality based on attack code and execution results. If either criterion is not met, we analyze the trial results and revise the relevant fields in $\mathcal{L}$. We then rerun \tool{} with the updated specification and repeat the checks. Once both criteria are satisfied, we retain $\mathcal{L}$ for subsequent dataset-construction runs for the same lineage.

\subsection{Pipeline Architecture of LineageForge}
\label{sec:overview}

We describe the pipeline architecture of \tool{} with the IGD and EGCC modules. Their underlying principles are detailed in Section~\ref{sec:discovery} and Section~\ref{sec:construction}, respectively.

\subsubsection{Compatible contract discovery with IGD}
As shown in Fig.~\ref{fig:pipeline}B, we first construct a script called \textit{Miner} to perform preliminary screening of contracts using Application Binary Interface (ABI) filtering followed by function-level filtering. We then perform a refinement operation to further screen these candidates and establish evidence bindings between the vulnerability requirements and contract source code. Through source analysis and evidence matching, refinement determines whether each candidate meets the requirements for $S_1$ or can be modified to meet them under the permitted-construction conditions in $\mathcal L$. The resulting evidence bindings are retained in $B_{C,L}$ for subsequent construction. Candidates that fail these checks are discarded. For each retained contract, refinement prepares a compilation workspace $W_0$ from the source files and compiler settings collected during source analysis.
IGD organizes each retained compatible contract $S_0$, $W_0$, and the corresponding binding packet $B_{C,L}$ into a source package. This package is provided to the EGCC module for constructing $S_1$, $S_2$, and their corresponding PoCs.

\subsubsection{Contract construction and validation with EGCC}
Prior work has used LLMs to generate vulnerable contract variants guided by business-logic specifications~\cite{xplogen2024}. As shown in Fig.~\ref{fig:pipeline}C, EGCC uses the lineage specification $\mathcal{L}$ and binding packet $B_{C,L}$ to guide an LLM-based planner and executor in constructing $S_1$ and $S_2$. Quality gate~1 combines compilation tests with LLM-based semantic review to assess whether $S_1$ enforces the intended protection, $S_2$ exhibits the target vulnerability, and both contracts preserve the required business behavior. If the review identifies violations, EGCC returns the findings and their causes to the planner for plan revision.

With both contracts passing quality gate 1, EGCC performs adaptive PoC preparation to generate the supporting code required to execute attacks against $S_2$. Following the historical exploit migration approach~\cite{pocshift2025}, the LLM executor uses this code and $\mathcal{L}$ to adapt the historical attack into a PoC for $S_2$. Once the PoC executes successfully and all assertions hold, the executor derives a corresponding PoC for $S_1$, preserving equivalent attacker actions and relevant preconditions. Once both PoCs compile and their expected assertions are satisfied during execution, quality gate 2 employs an LLM-based reviewer to examine their attack code and execution results and assign quality labels. PoCs labeled strong, acceptable, or weak are accepted and archived with the contract pair and execution environments as a dataset record. PoCs labeled unacceptable are rejected. For repairable PoC issues, the reviewer provides feedback to guide PoC revision and revalidation. Issues requiring contract modifications are instead returned to the planner for construction-plan revision.


\section{Incident-Guided Discovery}
\label{sec:discovery}


In this section, we present IGD, which uses $\mathcal{L}$ to identify compatible contracts from the source-verified contract corpus $\mathcal{C}$. As illustrated in Algorithm~\ref{alg:igd}, the \textit{Miner} script first narrows down the candidates through ABI filtering and function-level filtering. We then perform a refinement operation to further analyze the retrieved source code against the requirements in $\mathcal{L}$. For each retained contract $S_0$, we prepare its compilation workspace $W_0$ and binding packet $B_{C,L}$. The resulting set $\mathcal{R}$ contains these materials and serves as the input to EGCC.

\begin{algorithm}[t]
\caption{Incident-Guided Discovery}
\label{alg:igd}
$\mathcal R\gets\varnothing$\;
\algstage{Miner Script}\;
$Q\gets\algfn{filterABI}(\mathcal C,\mathcal L)$\;
$Q\gets\algfn{functionLevelFilter}(Q,\mathcal L)$\;
\algstage{Refinement}\;
\For{each $(C,e)\in Q$}{
  $\mathcal E\gets\algfn{profileSource}(C,e)$\;
  $(\mathrm{bindings},\Delta)\gets\algfn{evidenceMatching}(\mathcal E,\mathcal L)$\;
  \lIf{$\mathrm{bindings}=\bot$}{\textbf{continue}}
  $W_0\gets\algfn{materializeWorkspace}(C,e,\mathcal E)$\;
  $S_0\gets C$\;
  $B_{C,L}\gets\algfn{pack}(\mathrm{bindings},\mathcal E,\Delta,\mathcal L)$\;
  $\mathcal R\gets\mathcal R\cup\{(S_0,W_0,B_{C,L})\}$\;
}
\Return{$\mathcal R$}\;
\end{algorithm}

\subsection{\textit{Miner} Script for Preliminary Screening}
\label{sec:miner}

The \textit{Miner} script combines ABI filtering and function-level filtering to identify contract entries that may support the business logic required for $S_1$. Sourcify\footnote{\url{https://docs.sourcify.dev/docs/repository/download-dataset/}} provides a large corpus of real-world contracts, making it costly to retrieve all source code for screening. We therefore first apply ABI filtering to the separately accessible function signatures to narrow down the candidates. Function-level filtering then examines the source code of the corresponding contracts and excludes entries that lack the required features.

\subsubsection{Filtering requirements}
For the business context of the target vulnerability, $\mathcal{L}$ specifies function-signature patterns for ABI filtering and source-code features for function-level filtering. ABI filtering first excludes common state-reading function entries based on their signatures, and then retains entries matching the signature patterns specified in $\mathcal{L}$. The source-code features are grouped into three categories: business operations, relevant inputs, and validation-related code. For each group $j$, $\mathcal{L}$ defines a feature set $\mathcal{F}_j$ and a minimum matching threshold $\tau_j$. Each feature $f\in\mathcal{F}_j$ may be represented by multiple source-code patterns. The specification also defines whether matching any one pattern is sufficient or whether multiple patterns must be matched jointly for the feature to count as present.

\subsubsection{Filtering procedure}
As shown in Algorithm~\ref{alg:igd}, $\algfn{filterABI}$ applies the signature conditions to obtain the candidate set $Q$, where each element $(C,e)$ represents a contract $C$ and a function entry $e$ identified by its signature. Then, $\algfn{functionLevelFilter}$ locates each entry in the corresponding source code and checks its declaration and body against the feature-group requirements. We define $\sigma_{\mathcal L}(C,e)$ to indicate whether entry $e$ in contract $C$ satisfies the minimum matching thresholds of all three feature groups. $\sigma_{\mathcal L}(C,e)$ is expressed as
\begin{equation}
\label{eq:miner-groups}
\sigma_{\mathcal L}(C,e)=
\bigwedge_{j=1}^{3}\bigg[\sum_{f\in\mathcal F_j}m_f(C,e)\ge\tau_j\bigg],
\end{equation}
where $m_f(C,e)=1$ if the source code of entry $e$ matches feature $f$, and $m_f(C,e)=0$ otherwise. The indicator satisfies $\sigma_{\mathcal L}(C,e)=1$ only if all three feature groups meet their respective thresholds. It is worth noting that entries $(C,e)$ that pass this filtering stage are retained in $Q$ and proceed to the refinement operation for evidence matching.

\subsection{Refinement for Further Screening and Evidence Binding}
\label{sec:refine}

We further perform the refinement operation to examine the candidates returned by the \textit{Miner} script, thereby identifying compatible contracts and establishing source-level evidence bindings for subsequent construction. Since the features identified by the \textit{Miner} script provide only preliminary indications of compatibility, we analyze the source code to determine whether each candidate can serve directly as $S_1$ or can be adapted into $S_1$ through permitted code changes. 

\subsubsection{Refinement requirements}
Based on the vulnerability roles, protection conditions, and affected business operations, $\mathcal{L}$ specifies two types of construction conditions. Conditions for direct use define the business logic and protection requirements that a contract should already satisfy to serve as $S_1$. Conditions for permitted construction define the required business foundation and the protection logic or components that can be added through code modifications. To evaluate these conditions, $\mathcal{L}$ defines evidence-matching rules for each requirement, specifying the necessary source-level facts and how their matches are combined. In addition, $\mathcal{L}$ specifies binding, anchor, and effect requirements that define the associations among role-to-source bindings, code modification anchors, and protected business operations.

\subsubsection{Source analysis and preliminary assessment}
In IGD, $\algfn{profileSource}$ analyzes each candidate entry $(C,e)\in Q$ by parsing its source code and incorporating available Slither~\cite{feist2019slither} analysis results. It extracts evidence of parameter usage, conditional checks, state updates, and business operations, together with their corresponding source elements and locations, to form $\mathcal E$. $\algfn{evidenceMatching}$ then applies the evidence-matching rules in $\mathcal L$ to evaluate each requirement against $\mathcal E$. It retains the source elements and locations that support successful matches. These matches establish bindings between vulnerability roles and source elements, identify protected business operations, and locate existing code modification anchors.

We denote each requirement evaluated through grouped checks by $u$. For each requirement $u$, the evidence-matching rules define $k_u$ groups, where $\mathcal A_{u,j}$ denotes the set of checks in group $j$. A check $h(\mathcal E)$ evaluates to true when its corresponding evidence condition is satisfied. We define $\operatorname{sat}(u,\mathcal E)$ as a Boolean indicator of whether the extracted evidence satisfies requirement $u$, given by
\begin{equation}
\label{eq:refine-match}
\operatorname{sat}(u,\mathcal E)
=\bigwedge_{j=1}^{k_u}
\bigg(\bigvee_{h\in\mathcal A_{u,j}}h(\mathcal E)\bigg).
\end{equation}
Within each group, satisfying at least one check is sufficient for the group to pass, while requirement $u$ is satisfied only when all groups pass. Based on these matching results, $\algfn{evidenceMatching}$ assesses whether contract $C$ meets the direct-use or permitted-construction conditions. If direct use is feasible, it sets $\Delta=\varnothing$. Otherwise, for permitted construction, it derives the construction requirements $\Delta$ from the matching results and construction rules in $\mathcal L$.

\subsubsection{Association checks and final decision}
Even when the preliminary assessment indicates that a contract supports direct use or permitted construction, the matched evidence may refer to unrelated source elements. Hence, $\algfn{evidenceMatching}$ checks whether the role-to-source bindings, modification anchors, and protected business operations jointly satisfy the association constraints specified in $\mathcal L$. This check ensures that the identified business operation is associated with the source element bound to the corresponding vulnerability role. If $C$ satisfies neither the direct-use nor the permitted-construction conditions, or fails the applicable association checks, $\algfn{evidenceMatching}$ returns $(\bot,\varnothing)$. Here, $\bot$ indicates matching failure. Otherwise, the function returns $(\mathrm{bindings},\Delta)$, where $\mathrm{bindings}$ contains the successfully matched evidence bindings. These results are then used to construct the binding packet $B_{C,L}$ for EGCC.

\subsection{Source Package Preparation}

To support subsequent construction, IGD retains each compatible contract as $S_0$ and prepares its compilation workspace $W_0$. In Algorithm~\ref{alg:igd}, $\algfn{materializeWorkspace}$ assembles the required source files and dependencies and applies the compiler settings specified in the contract metadata to create $W_0$. This workspace supports the compilation of $S_1$ and $S_2$. $\algfn{pack}$ organizes the construction requirements $\Delta$, matched evidence bindings, and their supporting evidence into a binding packet $B_{C,L}$. The packet maps construction requirements to specific source elements and records the permitted code modifications and business behaviors to preserve, thereby guiding subsequent construction. Finally, IGD collects each tuple $(S_0,W_0,B_{C,L})$ into the source package set $\mathcal R$ and passes these packages to EGCC for constructing $S_1$, $S_2$, and their corresponding PoCs.

\subsection{Time Complexity of Algorithm 1}

For Algorithm~\ref{alg:igd}, we fix $\mathcal L$ and the source-analysis settings. We define $n$ as the number of $(C,e)$ pairs returned by ABI filtering and $s$ as the maximum total source size per contract, including dependencies. Moreover, we define $T_0$ as the total cost of ABI filtering, source and metadata retrieval, and global sorting and selection of filtering results. The functions $t_m(s)$ and $t_r(s)$ bound the per-pair costs of function-level filtering and refinement, respectively. The refinement bound includes source analysis, evidence matching, and preparation of $W_0$ and $B_{C,L}$ for successful matches. Under sequential execution, the time complexity of Algorithm~\ref{alg:igd} is $T_{\mathrm{IGD}}=\mathcal{O}\!\left(T_0+n\,t_m(s)+|Q|\,t_r(s)\right)$, where $|Q|$ counts the pairs retained after function-level filtering.

\section{Evidence-Grounded Contract Construction}
\label{sec:construction}

In this section, we present EGCC, which uses the source packages produced by IGD to construct $S_1$, derive its vulnerable counterpart $S_2$, and generate their corresponding PoCs. Guided by $\mathcal{L}$ and the evidence in $B_{C,L}$, EGCC either adopts $S_0$ directly as $S_1$ or modifies it within the permitted scope.


\begin{algorithm}[t]
\caption{Contract Construction and Validation}
\label{alg:egcr}
Initialize $F\gets\varnothing$ and $r_s,r_p\gets0$\;
\algstage{Construction Planning}\;
$P\gets\algfn{plan}^{\scriptscriptstyle\mathrm{LLM}}(S_0,\mathcal L,B_{C,L},F)$\label{ln:plan}\;
\lIf{$P.\mathrm{decision}=\texttt{unsupported\_or\_ambiguous}$}{\Return{failure}}
\lIf{$P$ is invalid}{\Return{failure}}
\algstage{Construction and Compilation}\;
\For{$i=1,2$ in order}{
  \lIf{$i=1$ and $P$ specifies direct use}{$S_1\gets S_0$}
  \lElse{$S_i\gets\algfn{construct}^{\scriptscriptstyle\mathrm{LLM}}(S_{i-1},P,i)$}
  $W_i\gets\algfn{copyWorkspace}(W_{i-1})$\;
  $W_i\gets\algfn{replaceFiles}(W_i,S_i)$\;
  $(S_i,W_i,g)\gets\algfn{compileRepair}^{\scriptscriptstyle\mathrm{LLM}}(S_i,W_i,P,b_c)$\;
  \lIf{$\neg g.\mathrm{passed}$}{\Return{failure with $g$}}
}
\algstage{Source Review}\;
$g\gets\algfn{sourceReview}^{\scriptscriptstyle\mathrm{LLM}}(S_1,S_2,P,\mathcal L)$\;
\If{$\neg g.\mathrm{passed}$}{
  \If{$g.\mathrm{replan}\land r_s<b_s$}{
    $F\gets(P,g.\mathrm{feedback})$ and $r_s\gets r_s+1$\;
    \textbf{go to} line~\ref{ln:plan}\;
  }
  \Return{failure with $g$}\;
}
\algstage{PoC Construction and Validation}\;
$(g,q,\mathrm{artifacts})\gets$ Algorithm~\ref{alg:poc} with $P,S_1,S_2,W_1,W_2$\;
\lIf{$g.\mathrm{accepted}$}{\Return{$(\algfn{record}(S_1,S_2,\mathrm{artifacts}),q)$}}
\If{$g.\mathrm{replan}\land r_p<b_p$}{
  $F\gets(P,g.\mathrm{feedback})$ and $r_p\gets r_p+1$\;
  \textbf{go to} line~\ref{ln:plan}\;
}
\Return{failure with $g$}\;
\end{algorithm}


\subsection{Contract Construction and Validation}
\label{sec:routing}

Selected $S_0$ contracts may differ in Solidity versions, function structures, and variable names. To accommodate these differences, the LLM planner tailors a construction plan to each $S_0$, specifying how to obtain $S_1$ when necessary and how to derive $S_2$ from it. The LLM executor then applies the plan to the contract source code.

\subsubsection{Construction requirements}
The construction requirements in $\mathcal L$ specify the protection to be enforced in $S_1$, the intended failure in $S_2$, and the business behavior to be preserved. The binding packet $B_{C,L}$ provides contract-specific requirements $\Delta$ and permitted modifications identified during refinement, together with the source evidence linking them to $S_0$. Based on the modification anchors in $B_{C,L}$ and the permissions in $\mathcal L$, EGCC identifies the editable contract components, such as the entry function and local helper functions. When constructing $S_1$ requires additional protection or supporting components, $\Delta$ specifies the corresponding requirements. Guided by the requirements $\Delta$, the LLM planner determines the necessary modifications, including permitted additions and their locations.

\subsubsection{Planning and construction}
As shown in Algorithm~\ref{alg:egcr}, the LLM planner uses a fixed output template to generate a construction plan $P$ based on $S_0$, $\mathcal L$, $B_{C,L}$, and any available feedback $F$. The construction plan $P$ is expressed as
\begin{equation}
\label{eq:construction-plan}
P=\algfn{plan}^{\scriptscriptstyle\mathrm{LLM}}(S_0,\mathcal L,B_{C,L},F).
\end{equation}
Initially, $F=\varnothing$. When replanning is triggered, $F$ stores the previous construction plan and the corresponding review feedback. Before construction, EGCC checks whether $P$ follows the required output format and whether its proposed modifications fall within the permitted scope.  

The LLM executor derives $S_1$ from $S_0$ according to the construction plan $P$. If $P$ specifies direct use, $S_1=S_0$. Otherwise, the LLM executor applies $\algfn{construct}^{\scriptscriptstyle\mathrm{LLM}}$ to modify the selected components of $S_0$ and establish the required protection. Thus, $S_1$ is expressed as
\begin{equation}
\label{eq:protected-construction}
S_1=
\begin{cases}
S_0, & \text{for direct use},\\
\algfn{construct}^{\scriptscriptstyle\mathrm{LLM}}(S_0,P,1), & \text{for permitted construction}.
\end{cases}
\end{equation}
Once $S_1$ passes compilation, the executor invokes the same construction function on $S_1$, using the part of $P$ that specifies the intended failure in $S_2$. The stage argument distinguishes this invocation from the earlier construction of $S_1$. Accordingly, $S_2$ is given by
\begin{equation}
\label{eq:vulnerable-construction}
S_2=\algfn{construct}^{\scriptscriptstyle\mathrm{LLM}}(S_1,P,2).
\end{equation}

Both versions are compiled in workspaces derived from $W_0$. EGCC applies $S_1$ to a copy of $W_0$ to prepare $W_1$. For $S_2$, it applies the modified source to a copy of the successfully compiled $W_1$, producing $W_2$. If compilation fails, compiler diagnostics guide LLM-based source repair within $P$. The function $\algfn{compileRepair}^{\scriptscriptstyle\mathrm{LLM}}$ compiles each version and performs up to $b_c$ repair attempts, returning the updated source and workspace with the compilation result $g$, which is given by
\begin{equation}
\label{eq:compile-repair}
(S_i,W_i,g)=\algfn{compileRepair}^{\scriptscriptstyle\mathrm{LLM}}(S_i,W_i,P,b_c), \: i\in\{1,2\}.
\end{equation}

The constructed contracts are expected to satisfy the specified protection, failure, and preservation requirements. EGCC assesses these properties through quality gate 1. Successful compilation is a prerequisite for this review. Once both versions compile, EGCC invokes $\algfn{sourceReview}^{\scriptscriptstyle\mathrm{LLM}}(S_1,S_2,P,\mathcal L)$ to assess whether $S_1$ enforces the required protection, $S_2$ exhibits the specified failure, and both preserve the required business behavior, and the function $\algfn{sourceReview}^{\scriptscriptstyle\mathrm{LLM}}$ returns a review result $g$. If the review indicates that replanning is needed and the limit of $b_s$ additional rounds has not been reached, EGCC records the current plan $P$ and review feedback in $F$ and returns them to the LLM planner for another construction round. If the review remains unsuccessful after the allowed replanning rounds, EGCC terminates the current attempt.


\subsection{PoC Construction and Validation}
\label{sec:verification}

To verify that the targeted code changes introduce the selected vulnerability, EGCC uses executable PoCs to confirm that the exploit succeeds on $S_2$ but is blocked on $S_1$. The attack requirements in $\mathcal L$ specify the attacker capabilities, preconditions, and actions to preserve from the historical attack. The outcome requirements define the protected business operation and its observable effects. EGCC uses Foundry's Forge test runner\footnote{\url{https://github.com/foundry-rs/foundry}} to compile and execute the PoCs, starting with the PoC for $S_2$. Once all assertions for the expected exploit outcome are satisfied, the LLM executor derives and executes a corresponding PoC against $S_1$, preserving equivalent attacker actions and relevant preconditions.

\begin{algorithm}[t]
\caption{PoC Construction and Validation}
\label{alg:poc}
\algstage{Adaptive PoC Preparation}\;
$K\gets\algfn{preparePoC}(S_2,W_2,\mathcal L)$\;
$G\gets\algfn{guidance}(P,S_1,S_2,\mathcal L,K)$\;
$H_2\gets\algfn{generatePoC}^{\scriptscriptstyle\mathrm{LLM}}(G,S_2)$\;
$r_r\gets 0$\;
\algstage{PoC Execution}\;
$q\gets\mathrm{unacceptable}$ and $\mathrm{artifacts}\gets\varnothing$\label{ln:poc-execution}\;
\For{$i=2,1$ in order}{
  \lIf{$i=1$}{$H_1\gets\algfn{derivePoC}^{\scriptscriptstyle\mathrm{LLM}}(H_2,S_1,G)$}
  $(g,k)\gets(\algfn{run}(H_i,S_i,W_i),0)$\;
  \While{$\neg g.\mathrm{passed}\land k<b_h$}{
    $H_i\gets\algfn{repairPoC}^{\scriptscriptstyle\mathrm{LLM}}(H_i,G,g)$\;
    $(g,k)\gets(\algfn{run}(H_i,S_i,W_i),k+1)$\;
  }
  $\mathrm{artifacts}[i]\gets(H_i,g,\text{execution environment})$\;
  \If{$\neg g.\mathrm{passed}$}{
    $g\gets\algfn{failureReview}^{\scriptscriptstyle\mathrm{LLM}}(g,G,\mathrm{artifacts})$\;
    \Return{$g,q,\mathrm{artifacts}$}\;
  }
}
\algstage{Quality Gate 2}\;
$(g,q)\gets\algfn{checkReview}^{\scriptscriptstyle\mathrm{LLM}}(H_1,H_2,G,\mathrm{artifacts})$\;
\lIf{$g.\mathrm{accepted}$}{\Return{$g,q,\mathrm{artifacts}$}}
\If{$g.\mathrm{repairable}\land r_r<b_r$}{
  $H_2\gets\algfn{repairPoC}^{\scriptscriptstyle\mathrm{LLM}}(H_2,G,g.\mathrm{feedback})$\;
  $r_r\gets r_r+1$\;
  \textbf{go to} line~\ref{ln:poc-execution}\;
}
\Return{$g,q,\mathrm{artifacts}$}\;
\end{algorithm}

\subsubsection{PoC generation and execution}
To reduce the burden of adapting the setup to each contract, adaptive PoC preparation uses source analysis of $S_2$ and predefined procedures to generate supporting code. These procedures select a deployment strategy, such as direct or proxy deployment, based on the initialization and proxy patterns in $S_2$, and generate mocks from available dependency definitions where applicable. This allows the LLM executor to focus on adapting the attack for $S_{2,F}$ to $S_2$. $\algfn{preparePoC}$ produces this supporting code $K$. EGCC combines $K$ with the construction plan, contract sources, and attack requirements to form the PoC generation context $G$, which is expressed as
\begin{equation}
\label{eq:poc-guidance}
G=\algfn{guidance}(P,S_1,S_2,\mathcal L,K).
\end{equation}
Using $G$, EGCC first generates the PoC $H_2$, given by
\begin{equation}
\label{eq:poc-vulnerable}
H_2=\algfn{generatePoC}^{\scriptscriptstyle\mathrm{LLM}}(G,S_2).
\end{equation}
The generated PoC prepares the required execution state, performs the attacker calls, and checks the expected effects through assertions. Forge compiles and executes the PoC $H_2$.

After $H_2$ compiles and its execution satisfies all assertions for the expected exploit outcome, the LLM executor derives the corresponding PoC $H_1$, which is expressed as
\begin{equation}
\label{eq:poc-protected}
H_1=\algfn{derivePoC}^{\scriptscriptstyle\mathrm{LLM}}(H_2,S_1,G).
\end{equation}
The derivation aims to preserve the attacker actions and relevant preconditions, including the caller, attack entry, arguments, and payment, while adapting the target contract and outcome assertions to $S_1$. EGCC then executes $H_1$ to check that the corresponding protection blocks the attack. For each version, $\algfn{run}(H_i,S_i,W_i)$ returns the compilation and execution results in $g$. Compilation failures or failed assertions guide $\algfn{repairPoC}^{\scriptscriptstyle\mathrm{LLM}}(H_i,G,g)$, and the revised PoC is re-executed. The counter $k$ tracks the repair attempts, with at most $b_h$ attempts allowed per version in each execution round. If $H_i$ still fails after these attempts, $\algfn{failureReview}^{\scriptscriptstyle\mathrm{LLM}}$ analyzes the failure and determines whether contract modifications are needed.

\subsubsection{Execution review and feedback}
\label{sec:quality}
Passing all assertions alone does not establish that the PoCs demonstrate the selected vulnerability. Once both PoCs compile and their expected assertions hold, quality gate~2 invokes $\algfn{checkReview}^{\scriptscriptstyle\mathrm{LLM}}$ to examine their attack code and execution logs. The reviewer assesses whether the two PoCs preserve equivalent attacker actions and relevant preconditions from the historical attack reproduced against $S_{2,F}$. It also checks whether the execution evidence links the successful exploit in $S_2$ and its blocking in $S_1$ to the specified failure and protection, respectively. Based on these assessments, the function returns an acceptance decision $g.\mathrm{accepted}$ and a quality label $q$ reflecting the strength of the execution evidence. If the PoCs are rejected, $g.\mathrm{accepted}=\mathrm{false}$, $q=\mathrm{unacceptable}$, and $g.\mathrm{feedback}$ records the identified issues and suggested corrections.

\begin{table*}[t]
\centering
\caption{Comparison of \tool{} and existing datasets in terms of data provenance, execution evidence, and construction capabilities.}
\label{tab:related-capabilities}
\papertablestyle
\setlength{\tabcolsep}{3pt}
\makeatletter
\def\relatedheaderbackground{\noalign{%
  \begingroup\color{gray!9}%
  \hrule height\dimexpr\ht\@arstrutbox+\dp\@arstrutbox\relax
  \vskip-\dimexpr\ht\@arstrutbox+\dp\@arstrutbox\relax
  \endgroup}}
\makeatother
\begin{tabular*}{\textwidth}{@{\extracolsep{\fill}}ll*{10}{c}@{}}
\toprule[1pt]
\relatedheaderbackground
\multicolumn{2}{c}{\textbf{Existing Datasets}} &
\cite{smartbugs2020} & \cite{dappscan2023} & \cite{forge2026} &
\cite{cyberchainbench2026} & \cite{evmbench2026} & \cite{solidifi2020} &
\cite{xplogen2024} & \cite{poco2025} & \cite{pocshift2025} & \textbf{\tool{}} \\
\midrule
\multirow{2}{*}{\textbf{Data provenance}}
 & Real-world contract code & $\checkmark$ & $\checkmark$ & $\checkmark$ & $\checkmark$ & $\checkmark$ & $\checkmark$ & & $\checkmark$ & $\checkmark$ & \textcolor{red}{$\checkmark$} \\
 & Historical incident cases & $\checkmark$ & & & $\checkmark$ & & & & & $\checkmark$ & \textcolor{red}{$\checkmark$} \\
\midrule
\multirow{2}{*}{\shortstack{\textbf{Execution evidence}}}
 & Executed exploits & & & & $\checkmark$ & $\checkmark$ & $\checkmark$ & $\checkmark$ & $\checkmark$ & $\checkmark$ & \textcolor{red}{$\checkmark$} \\
 & Vulnerable and protected execution & & & & $\checkmark$ & $\checkmark$ & & & $\checkmark$ & & \textcolor{red}{$\checkmark$} \\
\midrule
\multirow{2}{*}{\textbf{Construction}}
 & New vulnerable implementations & & & & & & $\checkmark$ & $\checkmark$ & & & \textcolor{red}{$\checkmark$} \\
 & Additional target matching & & & & & & & & & $\checkmark$ & \textcolor{red}{$\checkmark$} \\
\bottomrule[1pt]
\end{tabular*}
\end{table*}

The quality label $q$ takes one of four values: strong, acceptable, weak, or unacceptable. A strong label indicates that the attack actions, entry, roles, preconditions, and protected business effect are consistent with the historical attack reproduced against $S_{2,F}$. It also requires the specified failure and protection to explain the observed difference between $S_2$ and $S_1$. An acceptable label allows a simplified setup or business context, provided that the essential attack sequence and business effect are preserved without bypassing required preconditions or fabricating the effect. A weak label indicates that the observed difference relies on shortcuts or incomplete attack evidence. The first three labels correspond to accepted results, for which $g.\mathrm{accepted}=\mathrm{true}$. The unacceptable label indicates that the evidence fails to demonstrate the selected vulnerability and is assigned to rejected results.

For results rejected by quality gate~2, $g.\mathrm{feedback}$ records the identified issues and suggested revisions. If the issues are repairable at the PoC level, $g.\mathrm{repairable}$ is set to true. Within a limit of $b_r$ revisions per call to Algorithm~\ref{alg:poc}, EGCC invokes $\algfn{repairPoC}^{\scriptscriptstyle\mathrm{LLM}}(H_2,G,g.\mathrm{feedback})$ to revise $H_2$. EGCC then recompiles and executes the revised $H_2$. Once its expected exploit assertions hold, EGCC derives, compiles, and executes the corresponding $H_1$. Quality gate~2 reviews the updated PoC code and execution logs once both PoCs satisfy their expected assertions. If the review indicates that contract modifications are needed, $g.\mathrm{replan}$ is set to true, and $g.\mathrm{feedback}$ specifies the required changes. Algorithm~\ref{alg:poc} returns this feedback to Algorithm~\ref{alg:egcr}. Within a limit of $b_p$ additional replanning rounds per call to Algorithm~\ref{alg:egcr}, EGCC records the current plan $P$ and review feedback in $F$ and passes $F$ to the LLM planner to repeat contract construction. If neither PoC repair nor contract replanning can resolve the identified issues within their respective limits, the current attempt is terminated. Upon acceptance, EGCC archives $S_1$, $S_2$, their corresponding PoCs, and execution environments as a dataset record, together with the assigned quality label.

\subsection{Time Complexity of Algorithms 2 and 3}

For Algorithms~\ref{alg:egcr} and~\ref{alg:poc}, we fix the lineage specification, model, and tool settings. We define $t_L$, $t_C$, $t_F$, and $t_A$ as the cost bounds for one LLM operation, contract compilation, PoC compilation and execution, and other local operations, respectively. The LLM cost includes internal response retries. We further define $t_P$ as the cost bound for adaptive PoC and context preparation, including source analysis and supporting-code generation. These bounds depend on the maximum sizes of the inputs and intermediate artifacts. 

Within one call to Algorithm~\ref{alg:poc}, preparation occurs once, followed by at most $1+b_r$ execution rounds. Each round allows up to $1+b_h$ attempts per version. Thus, the time complexity of Algorithm~\ref{alg:poc} is $T_{\mathrm{PoC}}=\mathcal{O}(t_P+{(1+b_r)}{(1+b_h)}{(t_L+t_F+t_A)})$. The preparation cost $t_P$ is incurred before the loop of review and revision, while the remaining term accounts for generation, execution, repair, and review across both versions. Algorithm~\ref{alg:egcr} allows at most $1+b_s+b_p$ construction rounds for each source package. Each round includes up to $1+b_c$ compilation attempts per version and at most one invocation of Algorithm~\ref{alg:poc}. Hence, the time complexity of Algorithm~\ref{alg:egcr} is $T_{\mathrm{EGCC}}=\mathcal{O}({(1+b_s+b_p)}[{(1+b_c)}{(t_L+t_C+t_A)}+T_{\mathrm{PoC}}])$, accounting for planning, contract construction, compilation, source review, and PoC validation.

\section{Experimental Evaluation}
\label{sec:evaluation}

\edef\evaluationtopnumber{\number\value{topnumber}}
\setcounter{topnumber}{1}


\subsection{Comparison with Existing Datasets}

\begin{figure}[t]
  \centering
  \includegraphics[width=0.8\columnwidth]{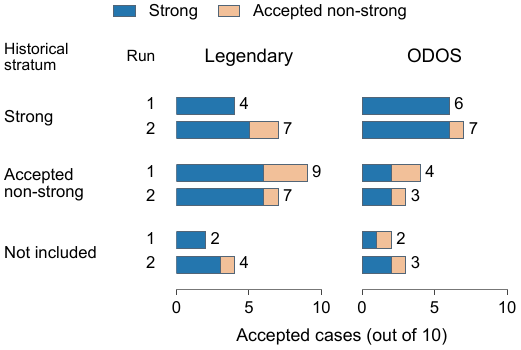}
  \caption{Construction outcomes across historical groups in the fixed $60$-case challenge. Bars count newly accepted records, colored by new pipeline grades.}
  \label{fig:rq1-construction}
\par\vspace{\floatsep}
  \includegraphics[width=0.8\columnwidth]{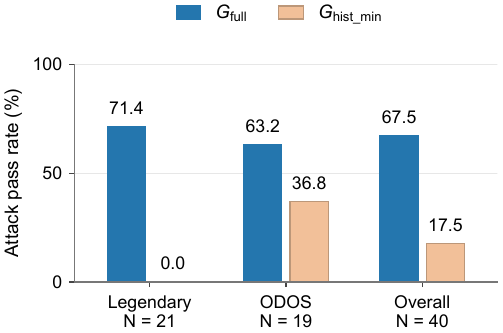}
  \caption{Attack-generation pass rates on the same $40$ contract pairs under the full-context and minimal-history configurations.}
  \label{fig:rq1-attack}
\end{figure}

\begin{table}[t]
  \hypersetup{urlcolor=magenta}
  \caption{Public provenance of the seven incident lineages. Month is the incident month in UTC.}
  \label{tab:incident-lineages}
\papertablestyle
  \centering
  
  \setlength{\tabcolsep}{2pt}
  \begin{tabular}{m{.16\columnwidth}|m{.13\columnwidth}m{.14\columnwidth}m{.24\columnwidth}m{\dimexpr.33\columnwidth-10\tabcolsep-\arrayrulewidth\relax}}
    \toprule[1pt]
    \rowcolor{gray!9}
    \textbf{Lineage} & \textbf{Month} & \textbf{Chain} &
    \textbf{Incident entry} & \textbf{Attack evidence} \\
    \hline
    BCBC\textsuperscript{\href{https://bscscan.com/address/0xbCbCb0e7E28414e084c4a40C1cCC30B75629a7DE\#code}{\ensuremath{\mathrm{a}}}} &
    2024-04 & BSC & \texttt{setReward} &
    \href{https://app.blocksec.com/explorer/tx/bsc/0xe15d6f7fa891c2626819209edf2d5ded6948310eaada067b400062aa022ce718}{\texttt{tx 0xe15d...e718}} \\
    ByToken\textsuperscript{\href{https://bscscan.com/address/0x6f50cffEcd4e00EcF7E442774C08c089450B62Ca\#code}{\ensuremath{\mathrm{a}}}} &
    2026-06 & BSC & \texttt{triggerAutoBurn} &
    \href{https://bscscan.com/tx/0xe31c681eee764fb94b1b6bda3bbb0e4f25acb129c19040b9f58ad30541980979}{\texttt{tx 0xe31c...0979}} \\
    SorStaking\textsuperscript{\href{https://etherscan.io/address/0x5d16b8Ba2a9a4ECA6126635a6FFbF05b52727d50\#code}{\ensuremath{\mathrm{a}}}} &
    2025-01 & Ethereum & \texttt{withdraw} &
    \href{https://app.blocksec.com/explorer/tx/eth/0x6439d63cc57fb68a32ea8ffd8f02496e8abad67292be94904c0b47a4d14ce90d}{\texttt{tx 0x6439...e90d}} \\
    ODOS\textsuperscript{\href{https://basescan.org/address/0xB6333E994Fd02a9255E794C177EfBDEB1FE779C7\#code}{\ensuremath{\mathrm{a}}}} &
    2025-01 & Base & \texttt{isValidSigImpl} &
    \href{https://basescan.org/tx/0xd10faa5b33ddb501b1dc6430896c966048271f2510ff9ed681dd6d510c5df9f6}{\texttt{tx 0xd10f...f9f6}} \\
    MURE\textsuperscript{\href{https://etherscan.io/address/0x365083717eFB17F3895290BA38f20F568C7A4D8a\#code}{\ensuremath{\mathrm{a}}}} &
    2026-05 & Ethereum & \texttt{distribute} &
    \href{https://etherscan.io/tx/0xb83040361a0ec72fa2d06ad69493226518a5f8b5d96c19b400626248f9c5b798}{\texttt{tx 0xb830...b798}} \\
    Legendary\textsuperscript{\href{https://bscscan.com/address/0x92D60629FF5d53a0098B51E9b1D59546D1D8e5B6\#code}{\ensuremath{\mathrm{a}}}} &
    2026-05 & BSC & \texttt{cliamRewred} &
    \href{https://bscscan.com/tx/0x15c835c070672948b3487d35254bce96831bec9f5212f78b04e683fed74bf4a2}{\texttt{tx 0x15c8...f4a2}} \\
    GPU\textsuperscript{\href{https://bscscan.com/address/0xf51CBf9F8E089Ca48e454EB79731037a405972ce\#code}{\ensuremath{\mathrm{a}}}} &
    2024-05 & BSC & \texttt{transfer} &
    \href{https://app.blocksec.com/explorer/tx/bsc/0x2c0ada695a507d7a03f4f308f545c7db4847b2b2c82de79e702d655d8c95dadb}{\texttt{tx 0x2c0a...dadb}} \\
    \bottomrule[1pt]
  \end{tabular}
  \par\smallskip
  \begin{minipage}{\columnwidth}
  \scriptsize
  $^{\mathrm{a}}$Links to the incident contract containing the incident-critical entry used by the lineage.
  \end{minipage}
\end{table}

In Table~\ref{tab:related-capabilities}, we compare \tool{} with existing datasets in terms of data provenance, execution evidence, and construction capabilities. Existing annotation datasets and execution benchmarks~\cite{smartbugs2020,dappscan2023,forge2026,evmbench2026,cyberchainbench2026} primarily organize collected contracts, vulnerability findings, and historical attack cases. Datasets based on vulnerability injection and synthesis~\cite{solidifi2020,xplogen2024} expand the available samples using predefined vulnerability patterns or business specifications. Meanwhile, exploit generation and historical attack reuse~\cite{poco2025,pocshift2025} provide executable PoCs to demonstrate exploitability. Building on these capabilities, \tool{} combines incident-guided construction with execution-based verification to expand vulnerabilities in real-world attacks across compatible contract implementations. Specifically, it uses incident-derived requirements to identify compatible contracts and construct corresponding protected and vulnerable versions. For each $S_1/S_2$ pair, \tool{} adapts and executes corresponding PoCs under equivalent attacker actions and relevant preconditions. Combined with source-code and PoC reviews, the execution results verify whether the targeted code changes enable the selected exploit in $S_2$ while the corresponding protection blocks it in $S_1$. Each accepted record retains its connection to the historical attack and includes executable evidence and a corresponding protected contract, supporting the evaluation of vulnerability detection across different contract implementations.

\subsection{Construction Outcomes and Attack Generation}
\label{sec:construction-profile}

Table~\ref{tab:incident-lineages} summarizes the seven incident lineages. We employ DuckDB~\cite{raasveldt2019duckdb}, Slither~\cite{feist2019slither}, and Foundry for contract retrieval, source-code profiling, and differential execution, respectively. All LLM-based operations in EGCC use DeepSeek V4 Flash. We also use the same model for the fixed-case experiments, with a temperature of $0.2$ and a maximum output length of $40{,}960$ tokens.

\subsubsection{Production scale}
The campaign processes $1{,}970$ of $2{,}306$ refined candidates ($85.4\%$) before reaching the target corpus size, leaving $336$ unprocessed. Of the processed candidates, $1{,}059$ pass PoC validation, yielding a pass rate of $53.8\%$. The dataset contains $901$ records assigned a strong grade, as reported in Table~\ref{tab:construction-profile}.

\subsubsection{Fixed \texorpdfstring{$60$}{60}-case challenge}
We conduct two full-pipeline executions using the same $30$ Legendary cases and $30$ ODOS cases. For each lineage, the test set contains $10$ cases from each of three historical groups: pipeline-strong, accepted non-strong, and not included. The not-included group consists of contracts that passed refinement but were absent from the historical dataset. All cases are selected before execution, and their group labels are withheld from the generator. Figure~\ref{fig:rq1-construction} reports the number of newly accepted records in each group. Both executions produce accepted records from all three strata, including the not-included group, with $27$ and $31$ of the $60$ cases accepted, respectively. The comparable acceptance totals and consistent group coverage suggest relatively stable aggregate construction outcomes across these two executions.

\begin{table*}[!t]
\centering
\caption{Observed historical production funnel across seven lineages. The ABI-selected pool and scanned rows count candidate function signatures in compilation records. Retrieved candidates are those returned by the \textit{Miner} script after source screening and collection limits. Matched counts candidates passing refinement, while Constr. and PoC in count cases entering the construction and PoC stages, respectively. Subsequent columns report stage-record or case counts. Compiled pairs require both contract versions to compile. Passed and Strong reflect historical pipeline outcomes and grades.}
\label{tab:construction-profile}
\papertablestyle
\scriptsize

\setlength{\tabcolsep}{3.1pt}
\begin{tabularx}{\textwidth}{@{}l|*{9}{>{\centering\arraybackslash}X}@{}}
\toprule[1pt]
\rowcolor{gray!9}
\textbf{Lineage} & \textbf{ABI pool} & \textbf{Scanned} &
\textbf{Retrieved} & \textbf{Matched} & \textbf{Constr.} &
\textbf{Compiled} & \textbf{PoC in} & \textbf{Passed} & \textbf{Strong} \\
\hline
\rowcolor{tableHeaderOrange!20}
BCBC & $1{,}088{,}237$ & $2{,}198$ & $1{,}000$ & $259$ & $250$ & $205$ & $193$ & $181$ & $150$ \\
\rowcolor{tableHeaderOrange!5}
ByToken & $392{,}264$ & $2{,}647$ & $1{,}000$ & $347$ & $329$ & $174$ & $173$ & $165$ & $151$ \\
\rowcolor{tableHeaderOrange!20}
SorStaking & $556{,}708$ & $109{,}483$ & $1{,}400$ & $142$ & $78$ & $77$ & $76$ & $67$ & $51$ \\
\rowcolor{tableHeaderOrange!5}
ODOS & $63{,}560$ & $51{,}184$ & $3{,}628$ & $752$ & $739$ & $455$ & $449$ & $360$ & $298$ \\
\rowcolor{tableHeaderOrange!20}
MURE & $1{,}216{,}257$ & $1{,}211{,}257^{\ddagger}$ & $4{,}602$ & $221$ & $96$ & $63$ & $63$ & $53$ & $50$ \\
\rowcolor{tableHeaderOrange!5}
Legendary & $140{,}591$ & $11{,}505$ & $3{,}000$ & $452$ & $428$ & $288$ & $285$ & $183$ & $151$ \\
\rowcolor{tableHeaderOrange!20}
GPU & $3{,}315{,}861$ & $85{,}947$ & $750$ & $133$ & $50$ & $50$ & $50$ & $50$ & $50$ \\
\hline
\rowcolor{blue!5}
\textbf{Total} & $\mathbf{6{,}773{,}478}$ & $\mathbf{1{,}474{,}221}^{\ddagger}$ & $\mathbf{15{,}380}$ & $\mathbf{2{,}306}$ & $\mathbf{1{,}970}$ & $\mathbf{1{,}312}$ & $\mathbf{1{,}289}$ & $\mathbf{1{,}059}$ & $\mathbf{901}$ \\
\bottomrule[1pt]
\end{tabularx}
\par\smallskip
\begin{minipage}{\textwidth}
\scriptsize

$^{\ddagger}$MURE scanned rows are reconstructed from the recorded pool size, excluding $5{,}000$ unassessed rows, which are not screening failures. Scanned counts exclude preloading, and totals sum lineage counts without cross-lineage deduplication.
\end{minipage}
\end{table*}

\begin{table}[t]
    \caption{Model-based semantic audit of $196$ sampled pipeline-strong records, where attack, effect, and fidelity denote the numbers of records passing both reviews for attack equivalence, differential-effect validity, and mechanism fidelity, respectively.}
    \label{tab:rq2-semantic-audit}
\papertablestyle
    \centering
    \resizebox{\columnwidth}{!}{
    \begin{tabular}{l|ccccc}
    \toprule[1pt]
    \rowcolor{gray!9}
    \textbf{Lineage} & \textbf{Samples} & \textbf{Attack} &
    \textbf{Effect} & \textbf{Fidelity} & \textbf{Confirmed} \\
    \hline
    \rowcolor{tableHeaderOrange!20}
    BCBC       & $33$ & $33$ & $33$ & $33$ & $33$ \\
    \rowcolor{tableHeaderOrange!5}
    ByToken    & $33$ & $32$ & $32$ & $33$ & $32$ \\
    \rowcolor{tableHeaderOrange!20}
    SorStaking & $11$ & $11$ & $11$ & $11$ & $11$ \\
    \rowcolor{tableHeaderOrange!5}
    ODOS       & $64$ & $64$ & $64$ & $64$ & $64$ \\
    \rowcolor{tableHeaderOrange!20}
    MURE       & $11$ & $11$ & $11$ & $10$ & $10$ \\
    \rowcolor{tableHeaderOrange!5}
    Legendary  & $33$ & $24$ & $32$ & $30$ & $22$ \\
    \rowcolor{tableHeaderOrange!20}
    GPU        & $11$ & $10$ & $9$  & $11$ & $9$  \\
    \hline
    \rowcolor{blue!5}
    \textbf{Total} & $\mathbf{196}$ & $\mathbf{185}$ & $\mathbf{192}$ &
    $\mathbf{192}$ & $\mathbf{181}$ \\
    \rowcolor{blue!5}
    \textbf{Rate} & - & $\mathbf{94.4\%}$ & $\mathbf{98.0\%}$ &
    $\mathbf{98.0\%}$ & $\mathbf{92.3\%}$ \\
    \bottomrule[1pt]
    \end{tabular}
    }
\end{table}

\subsubsection{Attack-generation comparison}
We compare the full-context configuration $G_{\mathrm{full}}$ and the minimal-history configuration $G_{\mathrm{hist\_min}}$ using the same $40$ compiled contract pairs from the first full-pipeline run, including $21$ from Legendary and $19$ from ODOS. We reuse the PoC outcomes from that run for $G_{\mathrm{full}}$. Under $G_{\mathrm{full}}$, the generator receives the construction context and outcome requirements. Under $G_{\mathrm{hist\_min}}$, it receives only the historical incident attack for $S_2$ and the protected reference attack for $S_1$, without access to the construction plan or $S_1$ outcome requirements. A pair is considered to pass when both PoC harnesses pass their execution checks and the PoC review accepts the pair, as described in Section~\ref{sec:verification}. As shown in Fig.~\ref{fig:rq1-attack}, the full-context configuration passes $27$ of $40$ pairs, compared with $7$ for the minimal-history configuration. It also achieves higher pass rates in both lineages, resulting in an aggregate gap of $50.0$ percentage points. These results demonstrate that \tool{} can construct executable contract pairs across the evaluated lineages and that providing construction context and outcome requirements substantially improves attack-generation performance on the same fixed contract pairs.


\begin{figure}[t]
	\centering
	\subfigure[ODOS]
	{\includegraphics[width=0.23\textwidth]{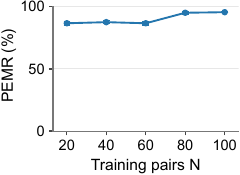}\label{ODOS}}
	\subfigure[Legendary]
	{\includegraphics[width=0.23\textwidth]{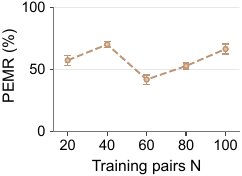}\label{Legendary}}
	\caption{Paired training on ODOS and Legendary. Points and whiskers show five-seed mean PEMR and sample standard deviation on fixed test sets.}\label{fig:rq3-training}
\end{figure}

\subsection{Semantic Fidelity}
\label{sec:rq2-fidelity}

\subsubsection{Experimental setup}
We sample $196$ of the $901$ pipeline-strong records using fixed-seed hash ordering within each lineage, retaining cases with missing evidence. Two independent review contexts of \texttt{gpt-5.6-luna} assess the same contract source, PoC and helper code, and execution records, with the prior strong grade disclosed. Both follow a shared review protocol and lineage-specific criteria covering attack equivalence, differential-effect validity, and mechanism fidelity. For Legendary, attack equivalence requires both PoCs to use the same business entry, vulnerability-relevant inputs, pre-attack state, and attack sequence. Differential-effect validity requires $S_1$ to block the unauthorized claim or preserve the protected state, while $S_2$ produces the target effect through the selected contract path under comparable outcome checks. Mechanism fidelity further requires $S_2$ to execute signer recovery and incorrectly accept a zero recovered signer against the authority state, enabling the protected claim or authorization effect. We accept target-contract state changes as evidence, including allowance and nonce consumption or observable mint and claim effects. We exclude cases that bypass signer recovery or rely on a valid attack signature, an unrelated revert in $S_1$, or effects produced by the test harness or mocks. A pair passes only when both reviews accept it under all three criteria.

\subsubsection{Numerical results}
Table~\ref{tab:rq2-semantic-audit} summarizes the semantic audit results for records assigned a strong label by the pipeline. We observe that $181$ of the $196$ sampled records pass all three criteria in both reviews, yielding a joint confirmation rate of $92.3\%$. In addition, attack equivalence has a lower confirmation rate of $94.4\%$, compared with $98.0\%$ for both differential-effect validity and mechanism fidelity. Overall, these results demonstrate that most pipeline-strong labels are supported by independent semantic reviews, providing further evidence of their credibility.

\subsection{Learnability of Contract Pairs}
\label{sec:rq3-training}

\begin{table}[t]
\centering
\caption{Train, Val., and Test denote the numbers of $S_1/S_2$ pairs, whereas Ext. neg. and Test tot. denote the numbers of individual test contracts. Test tot. is calculated as $2\times$ Test $+$ Ext. neg. The $22$ Ext. neg. for Legendary are a subset of the $45$ for ODOS and are not counted again in aggregate totals.}
\label{tab:rq3-data-splits}
\papertablestyle

\setlength{\tabcolsep}{4pt}
\begin{tabular}{p{\dimexpr.228\columnwidth-2\tabcolsep\relax}|*{3}{>{\centering\arraybackslash}p{\dimexpr.152\columnwidth-2\tabcolsep-.2\arrayrulewidth\relax}}
>{\centering\arraybackslash}p{\dimexpr.164\columnwidth-2\tabcolsep-.2\arrayrulewidth\relax}
>{\centering\arraybackslash}p{\dimexpr.152\columnwidth-2\tabcolsep-.2\arrayrulewidth\relax}}
\toprule[1pt]
\rowcolor{gray!9}
\textbf{Lineage} & \tablehead{\textbf{Train}} & \textbf{Val.} & \tablehead{\textbf{Test}} & \tablehead{\textbf{Ext. neg.}} & \tablehead{\textbf{Test tot.}} \\
\hline
\rowcolor{tableHeaderOrange!20}
ODOS & $208$ & $45$ & $45$ & $45$ & $135$ \\
\rowcolor{tableHeaderOrange!5}
Legendary & $104$ & $23$ & $22$ & $22$ & $66$ \\
\bottomrule[1pt]
\end{tabular}
\par\smallskip
\begin{minipage}{\columnwidth}
\footnotesize
\end{minipage}
\end{table}

\subsubsection{Data}
We use $298$ ODOS and $149$ Legendary contract pairs without additional filtering. Within each lineage, contract pairs and samples linked by case identifiers, source hashes, or target-file paths are kept together during splitting, as shown in Table~\ref{tab:rq3-data-splits}. We use DIVE samples whose eight vulnerability labels are all $0$ as external negatives~\cite{dive2026}. We exclude any overlap with the paired samples and select external negatives that approximate the token-length distributions of the paired training pools. External negatives for training and testing are drawn from separate groups formed using code and deployment metadata. Using fixed hash ordering, we select $45$ external negatives for ODOS and $22$ for Legendary, yielding a $1{:}1{:}1$ ratio of $S_1$, $S_2$, and external negatives in each test set. Within each lineage, all conditions, training sizes, and seeds use the same validation and test sets.

\subsubsection{Training setup}
For each lineage, we train a separate Multi-Layer Perceptron (MLP) from scratch so that no model is shared across lineages. For a given lineage, the inputs are complete Solidity files, labeled as target-negative for $S_1$ and target-positive for $S_2$. Before feature extraction, we tokenize each contract while ignoring comments and whitespace. The resulting tokens are then converted into a TF-IDF representation using case-sensitive token $1$--$3$-grams, at most $12{,}000$ features, and sublinear term frequency. To reduce dimensionality, we apply truncated Singular Value Decomposition (SVD) with seed $42$ and rank $\min(160,m-1,v-1)$ for $m$ training samples and $v$ features. The projected representation is then $\mathrm{L}_2$-normalized and zero-padded to $160$ dimensions. To avoid feature-fitting leakage, TF-IDF and truncated SVD are fitted only on the selected training samples of each condition. We then determine the input dimensionality by validating Binary Cross-Entropy (BCE) using the complete paired training pools and fixed validation sets. Among $128$, $160$, and $192$ dimensions, we choose the smallest whose score is within $0.005$ of the best score, where the score is the minimum validation BCE averaged over $5$ seeds and equally over both lineages. With the representation fixed, we train the MLP using Adam with learning rate $10^{-3}$, weight decay $10^{-4}$, and gradient clipping $1.0$ for $100$ epochs with batch size $32$. We select the checkpoint with the lowest validation BCE and, for prediction, classify a sample as vulnerable if the predicted probability is $\geq0.5$.

\begin{table}[t]
\centering
\caption{Each condition uses $200$ training examples with the same $100$ $S_2$ examples. Values are percentages (five-seed mean $\pm$ sample standard deviation). Arrows indicate whether higher ($\uparrow$) or lower ($\downarrow$) values are better.}
\label{tab:rq3-comparison}
\papertablestyle
\setlength{\tabcolsep}{2.5pt}
\begin{tabular}{>{\raggedright\arraybackslash}m{\dimexpr.19\columnwidth-2\tabcolsep\relax}
m{\dimexpr.29\columnwidth-2\tabcolsep-\arrayrulewidth\relax}|
*{2}{>{\centering\arraybackslash}m{\dimexpr.26\columnwidth-2\tabcolsep\relax}}}
\toprule[1pt]
\rowcolor{gray!10}
\textbf{Lineage} & \textbf{Metric} & \textbf{Paired} & \textbf{Unpaired} \\
\hline
\multirow{4}{*}{\textbf{ODOS}} 
  & \cellcolor{tableHeaderOrange!20}Macro-F1$_{\mathrm{all}}$ $\uparrow$ & \cellcolor{tableHeaderOrange!20}$95.37\pm0.44$ & \cellcolor{tableHeaderOrange!20}$68.13\pm1.15$ \\
  & \cellcolor{tableHeaderOrange!5}PEMR $\uparrow$ & \cellcolor{tableHeaderOrange!5}$95.56\pm0.00$ & \cellcolor{tableHeaderOrange!5}$4.44\pm3.51$ \\
  & \cellcolor{tableHeaderOrange!20}FPR$_{S1}$ $\downarrow$ & \cellcolor{tableHeaderOrange!20}$0.00\pm0.00$ & \cellcolor{tableHeaderOrange!20}$94.67\pm4.61$ \\
  & \cellcolor{tableHeaderOrange!5}FPR$_{\mathrm{safe}}$ $\downarrow$ & \cellcolor{tableHeaderOrange!5}$8.00\pm1.22$ & \cellcolor{tableHeaderOrange!5}$0.00\pm0.00$ \\
\hline
\multirow{4}{*}{\textbf{Legendary}} 
  & \cellcolor{tableHeaderOrange!20}Macro-F1$_{\mathrm{all}}$ $\uparrow$ & \cellcolor{tableHeaderOrange!20}$74.81\pm1.82$ & \cellcolor{tableHeaderOrange!20}$62.21\pm10.82$ \\
  & \cellcolor{tableHeaderOrange!5}PEMR $\uparrow$ & \cellcolor{tableHeaderOrange!5}$66.36\pm4.07$ & \cellcolor{tableHeaderOrange!5}$0.91\pm2.03$ \\
  & \cellcolor{tableHeaderOrange!20}FPR$_{S1}$ $\downarrow$ & \cellcolor{tableHeaderOrange!20}$18.18\pm3.21$ & \cellcolor{tableHeaderOrange!20}$98.18\pm2.49$ \\
  & \cellcolor{tableHeaderOrange!5}FPR$_{\mathrm{safe}}$ $\downarrow$ & \cellcolor{tableHeaderOrange!5}$39.09\pm2.49$ & \cellcolor{tableHeaderOrange!5}$12.73\pm28.46$ \\
\bottomrule[1pt]
\end{tabular}
\par\smallskip
\begin{minipage}{\columnwidth}
\footnotesize
\end{minipage}
\end{table}

\subsubsection{Evaluation metrics}
We evaluate performance using complementary metrics. Pairwise Exact-Match Rate (PEMR) is the proportion of test pairs for which both members are classified correctly, thereby capturing pair-level correctness. Macro-F1$_{\mathrm{all}}$ averages the two class F1 scores over the full test set and summarizes overall class-balanced performance. To characterize false-positive behavior, FPR$_{S1}$ and FPR$_{\mathrm{safe}}$ measure the false-positive rates on protected members and external negatives, respectively. For all metrics, we report means and sample standard deviations across five model seeds, $42$--$46$, with data selections held fixed across seeds.

\subsubsection{Learnability across training sizes}
In the paired setting, we use nested subsets of $N=20,40,60,80,100$ matched contract pairs, so that each subset provides $2N$ training examples. Figure~\ref{fig:rq3-training} shows that the resulting learning curves are non-monotonic. Specifically, as $N$ increases from $20$ to $100$ pairs, the mean PEMR increases from $86.67\%$ to $95.56\%$ on ODOS and from $57.27\%$ to $66.36\%$ on Legendary. ODOS maintains higher pairwise accuracy throughout, whereas Legendary drops from $70.00\%$ at $40$ pairs to $41.82\%$ at $60$ pairs before recovering. These results indicate that the constructed contract pairs support learning the vulnerable-protected distinction on held-out contracts in both tested lineages, although adding more training pairs does not consistently improve accuracy.

\subsubsection{Negative-example comparison}
At $N=100$, the unpaired condition replaces the $100$ matched $S_1$ examples with $100$ external negatives while retaining the same $100$ $S_2$ examples and the same training budget. As shown in Table~\ref{tab:rq3-comparison}, the paired condition substantially improves paired classification. Specifically, PEMR increases from $4.44\%$ to $95.56\%$ on ODOS and from $0.91\%$ to $66.36\%$ on Legendary. Concurrently, FPR$_{S1}$ decreases and full-test Macro-F1$_{\mathrm{all}}$ increases in both lineages. However, the mean FPR on external negatives increases from $0.00\%$ to $8.00\%$ on ODOS and from $12.73\%$ to $39.09\%$ on Legendary. These results indicate that matched $S_1$ negatives better support vulnerable-protected classification than external negatives do.

\setcounter{topnumber}{\evaluationtopnumber}

\section{Conclusion}
\label{sec:conclusion}

In this paper, we proposed \tool{}, an LLM-based framework for constructing verifiable SCV datasets at scale from vulnerabilities observed in real-world attacks. Within \tool{}, we first designed the IGD module to identify a predefined number of compatible contracts from a source-verified contract corpus. To construct and validate vulnerability instances using $\mathcal{L}$ and the source packages produced by IGD, we further proposed the EGCC module, which guides LLMs to construct a protected contract and to derive its vulnerable counterpart through targeted code changes. Based on seven real-world incidents, \tool{} produces $1{,}059$ validated instances, each pairing a protected contract with its vulnerable counterpart and including executable PoC attacks and the execution environments required for reproducible verification. For future work, we will extend \tool{} to additional vulnerability mechanisms and further automate incident analysis and lineage specification. We will also standardize construction requirements and strengthen validation procedures to improve the consistency and verifiability of \tool{}.



\bibliographystyle{IEEEtran}
\bibliography{references}

\end{document}